# Experimental study of pedestrian flow mixed with wheelchair users through funnel-shaped bottlenecks

Hongliang Pan, Jun Zhang[*], Weiguo Song

State Key Laboratory of Fire Science, University of Science and Technology of China, Hefei 230027, People's Republic of China

**Abstract**

With the increase of the elderly and disabled in the world, the characteristics of pedestrian flow mixed with wheelchair users has being paid more and more attention. In this study, experiments in funnel-shaped bottlenecks were performed to study the impact of bottleneck shape and the ratio of wheelchair users on the crowd dynamics. It is found that the increase of wheelchairs in the crowd obviously leads to worse moving efficiency and congestion from the escape time, time-space relationship and time headway. Under low mixing ratios (<2.35%), less congestion occurred in the 45° bottleneck among the four tested angles (0°, 15°, 30°, 45°). The average speeds of wheelchair users is the fastest in 45° bottleneck (0.310±0.097m/s) until the mixing ratio arrives at 7.05%. However, the advantage of the angle disappears when the mixing ratio gets higher. The findings in this study is meaningful for the guidance of pedestrian evacuation through bottlenecks with the presence of wheelchair users.

## 1. Introduction

Nowadays, the number of the elderly and the disabled is increasing due to the aging of population all over the world. According to the second national sampling survey on disabled people, 6.34% (82.96 million) of people in China have various physical impairments [1]. In addition, the number of people with mobility impaired has increased significantly compared with the first national sampling survey in China [2]. Therefore, the travel safety of this group is paid more and more attentions. For the purpose of safety evacuation and barrier-free facilities design, it is of great significance to study the characteristics of pedestrian flows including wheelchair users.

Many experimental studies on heterogeneous pedestrian flows consider limited characteristics such as age, gender, emotion (competition, selfish attitude) [3-6] or disabilities including visual, physical and other types of impairments [7-11]. Recently, some studies focus on the characteristics of the pedestrian flows by considering physical disabilities, especially the wheelchair users [12-16]. Sharifi et al. summarized the individual speeds of wheelchair users under different congestion levels and environments [12, 13]. It is found that the maximum walking speed in passageway can be reached at low congestion levels. The average speed of wheelchair users with an aid in the experimental scenarios of passageways is 1.06m/s, which is smaller than able-bodied people and barely influenced by gender [14]. Miyazaki et al. [15] studied the overtaking behavior of able-bodied participants in the flow including one wheelchair user experimentally. According to the experiments by Geoerg et al., the average walking speeds of pedestrians are 0.35±0.23m/s (participants at the average age of 37±16 mixed with wheelchair users) and 0.37±0.23m/s (participants at the average age of 32±16 without wheelchair users) in a symmetrical bottleneck with the width of 0.9m and length of 2.4m [16].

[*] Corresponding author: junz@ustc.edu.cn

To gain more insight into the traffic efficiency of pedestrian flows including wheelchair users, Shimada et al. [17] analyzed the impact of different mixing ratios of wheelchair users on pedestrian flow. The experimental results show that the traffic efficiency decreases with the increase of the mixing ratio while the width of bottleneck has less impact. Besides, the bottleneck width shows linear relations to the flow [18-21] and the evacuation efficiency of the bottlenecks is higher in the circumstances of the funnel angle between 46° and 65° [22, 23]. However, these studies are less concentrated on the influence of bottleneck of angles between 0° and 45° and wheelchair users. Based on the time-space diagrams of pedestrian flow with different populations (namely with and without wheelchair users), Geoerg et al. [16, 24] found that the jams occur more frequently in the presence of wheelchair users. Another effective factor utilized to reflect the congestion of the crowd is time headway, which has been adopted both in animal experiments and pedestrian flows. Garcimartín et al. [31] calculated the time headway of the sheep herding and adopted power law distribution to quantify the clogging. Referring to pedestrian flows, less congestion is found in the scenario where the exit was at the corner instead of the middle [27]. However, it is not clear how the congestion is at different mixing ratios of wheelchair users or bottleneck angles.

Based on these considerations, we conducted a series of controlled experiments to explore the effect of the ratios of wheelchair users on pedestrian flows through funnel shaped bottlenecks with different angles. The paper is organized as follows: Section 1 introduces the background and the state of the art on pedestrian flows including wheelchair users through bottlenecks. Section 2 describes the setup of the controlled pedestrian experiment. Section 3 analyzes the experimental data and discusses the result. Finally, section 4 is the conclusions of this study and recommendations of the future research.

**2. Experiment**

The purpose of the experiment is to study the influence of ratios of wheelchair users and angles of the funnel-shaped bottlenecks on pedestrian flows. The design of the experiment was inspired from the previous studies [17, 22, 23], which explored the evacuation efficiency by adding a funnel-shaped buffer zone in front of a bottleneck [22, 23] and changing the mixing rate of wheelchair users [17]. Two variables are considered in the experiment setup:

(1) Mixing ratio: namely the mixing ratio of wheelchair users and assistants. It is defined as the ratio of the number of wheelchair users and assistants to the number of the total pedestrians.

(2) Bottleneck angle: the change of the angles in the outer margin of the bottleneck entrance, which is shown in Fig.1.

2.1 Experimental setup

As shown in Fig. 1(a), the experiment was conducted on November, 10, 2018 in front of a teaching building where the space was large enough and the ground is flat. The weather condition with appropriate temperature and humidity was comfortable for participants. Fig. 1(b) shows the sketch of the experiment setup. The scenario of the experiment is a bottleneck with a dimension of 4.8m ×3.5m × 1.8m (length, width, height) surrounded by the boards with a height of 1.8m and width of 0.8m. In this study the angle of the bottleneck is defined as the angle between the dotted line and the vertical line of the bottleneck entrance (namely θ in Fig 1(b)) and set to 0°, 15°, 30°, 45° separately by moving the boards around the bottleneck entrance in the experiment.

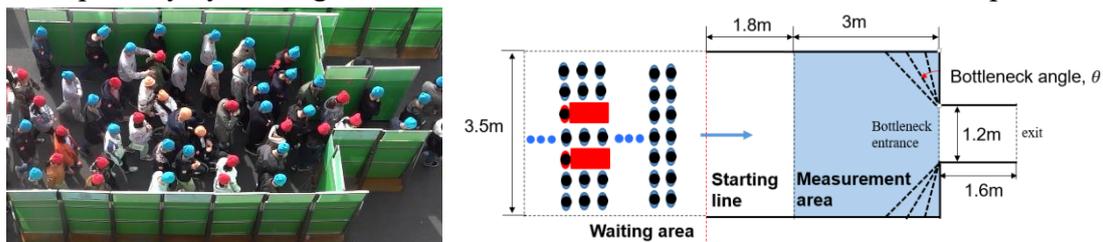

**Fig. 1** (a) A snapshot of the experiment (b) Sketch of the experiment setup. The scenario of bottleneck is consisted of two parts: one is from the starting line to the bottleneck entrance with a dimension of 4.8m (length) ×3.5m (width) and another is from the bottleneck entrance to the exit with the size of 1.6m (length) ×1.2m (width). The blue shaded area with the length of 3m and width of 3.5m shows the measurement area adopted in our following analysis. Only the front and middle of the team are shown in the waiting area, while the others are represented by the blue dots. The schematic diagram of two wheelchair users are shown here. The angle between the dotted line and the vertical line of the bottleneck entrance is defined as bottleneck angle θ and set to 0°, 15°, 30°, 45° separately by moving the boards around the bottleneck entrance in the experiment.

2.2 Participants

Totally 85 able-bodied participants (40 females and 45 males) with an average height of 169±7.6cm and average age of 22±2.1 years were recruited from universities. Six were selected to mimic wheelchair users and their assistants. The three assistants were trained to push the wheelchairs and to avoid collisions with others before the experiment. The wheelchairs used in this experiment was Hubang manual wheelchair HBL33, which are made by Hubang medical devices company in Shanghai and compliance with the relevant national standards. The specifications of the wheelchairs are shown in Fig.2.

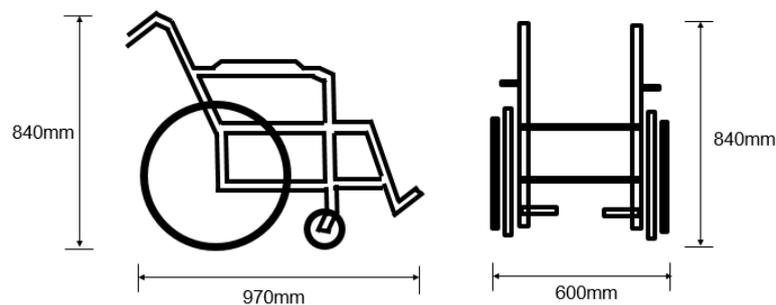

**Fig. 2** The specifications of Hubang manual wheelchair HBL33

2.3 Experimental procedure

The experiment scenarios include four symmetrical bottlenecks with different angles (0°, 15°, 30° and 45°). Meanwhile, 4 trials with different number of wheelchair users (0, 1, 2, 3) were performed in each bottleneck scenario. Table 1 shows the details of the experimental scenarios. Considering the width of the waiting area and the discipline management of the procedure, the participants were asked to stand in seven columns facing the bottleneck in the waiting area before the start of each trial. The wheelchair users and their assistants pushing the wheelchairs were spaced evenly in the middle of the whole team given that the location of the wheelchairs is not researched in this study. The crowd stood in the waiting area as shown in Fig. 1(b) and began to move towards the exit in normal upon hearing the signal of start. After finishing several trials, the participants were allowed to have a rest for 10 minutes.

Table 1 Parameters of the experiment scenarios

| Index | Scenario | Angle of bottleneck | Mixing ratio of wheelchair users and assistants |
|---|---|---|---|
| 1 | BW-0-0 | 0° | 0 |
| 2 | BW-0-1 | 0° | 2.35% |
| 3 | BW-0-2 | 0° | 4.70% |
| 4 | BW-0-3 | 0° | 7.05% |
| 5 | BW-15-0 | 15° | 0 |
| 6 | BW-15-1 | 15° | 2.35% |
| 7 | BW-15-2 | 15° | 4.70% |
| 8 | BW-15-3 | 15° | 7.05% |
| 9 | BW-30-0 | 30° | 0 |
| 10 | BW-30-1 | 30° | 2.35% |
| 11 | BW-30-2 | 30° | 4.70% |
| 12 | BW-30-3 | 30° | 7.05% |
| 13 | BW-45-0 | 45° | 0 |
| 14 | BW-45-1 | 45° | 2.35% |
| 15 | BW-45-2 | 45° | 4.70% |
| 16 | BW-45-3 | 45° | 7.05% |

2.4 Data extraction

The whole experiments were recorded by high definition digital cameras installed vertically 12m from the ground with the frame rate of 25fps. The trajectories of the participants were automatically extracted from the video footages using the pedestrian tracking software PeTrack [32]. The trajectories were manually examined and corrected after the automatic extraction from the software in order to get precise tracking data, which are shown in Fig 3. Afterwards, the variables on pedestrian movement including speed, escape time can be calculated from those trajectories.

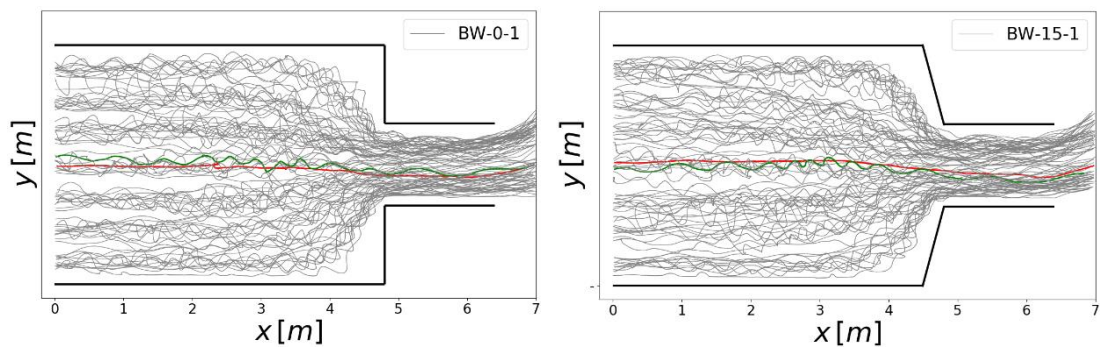

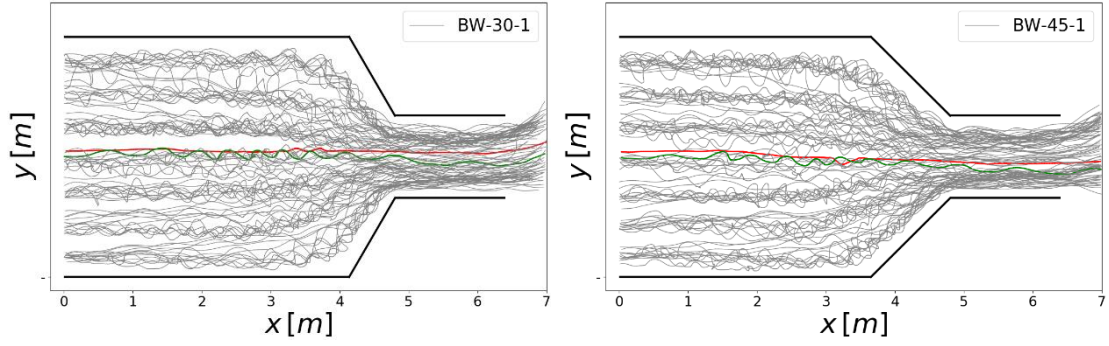

**Fig. 3** The trajectories of the participants in the bottleneck experiment with mixing ratio of 2.35% (namely one wheelchair user and one assistant). Those in other scenarios are shown in Appendix A. The grey lines represent the trajectories of the able-bodied participants, while the red one means the trajectories of the wheelchair user and the green one means that of the assistant.

## 3. Results and discussions

The movement of pedestrians in the bottleneck is considered as low velocity and clogging [33]. In order to gain insight into the characteristics of motion and congestion, we adopt the following factors including walking speed, escape time, time-space relation and time headway. These factors can reflect the sporting abilities, efficiency of the movement in crowd, stop and go waves during the motion and the congestion of crowd in the bottleneck entrance separately.

It is worth noticed that the last 7 participants who stood in the last row of the crowd fell behind and wandered around the bottleneck. Therefore, the data from them have been ruled out in the following analysis, which means that only 91.7% (78 participants) of the whole crowd are considered. The measurement area in the following analysis is the area of 3m ×3.5m in front of the exit, which is shown in blue shaded area in Fig. 1(b).

### 3.2 Speed

In order to reflect the physical abilities of participants in the bottlenecks, the instantaneous speed of each participant $v_i(t)$ is adopted and calculated by the values of the position $p_i$ in a small time interval $\Delta t'$ around $t$, which is defined as:

$$v_i(t) = \frac{p_i\left(t+\frac{\Delta t'}{2}\right)-p_i\left(t-\frac{\Delta t'}{2}\right)}{\Delta t'} \tag{1}$$

Based on the equation (1), the average speed of each participant in the measurement area can be obtained by calculating the ratio between the sum of the instantaneous speeds passing through the measurement area and the number of them. As shown in Fig. 4(a), the ordinate of each point represents the average speed of each participant in the measurement area based on the above method and the abscissa represents the sequence each participant arriving at the bottleneck entrance. The earlier a participant arrives at the bottleneck entrance, the smaller the abscissa of the point is. The sequence, namely the abscissa, of the first one who arrived at the bottleneck entrance is set to 1. The evolution of the speeds can be divided into two stages. In the first stage it is found that the average speeds of the participants who arrived early decline quickly. While the speeds enter the stationary stage as the sequence of arrival becomes larger (nearly larger than 40). It is indicated that these pedestrians who arrived at the bottleneck entrance earlier have a larger average speed in the measurement area. The speeds of those who arrived later get smaller compared to the pedestrians arriving earlier. The reasons for this phenomenon are as follows: the speed of the participants who arrived at the bottleneck entrance earlier is larger because there are no pedestrians standing in front of them and their movements are close to unimpeded movement especially for the participants standing at the head of the team. Afterwards the speed of the participants get smaller because those participants who arrived earlier than them lead to congestion at the bottleneck. Meanwhile, it is obvious that the speeds of wheelchair users (red scatters in Fig. 4(a)) are smaller than that of the able-bodied participants whose sequence of arrival is similar to wheelchair users (the black line in Fig. 4(a)). Moreover, the differences of the speeds of wheelchair users among different scenarios are studied by calculating the average speed and the results are shown in Fig. 4(b). Table 2 lists the average speeds of wheelchair users in different scenarios and the p value (by T-test) between angle 0° and other three angles at different mixing ratios. We find that the average speeds of wheelchair users decrease with the increase of the mixing ratios. The average speeds of wheelchair users in the measurement area is the fastest in angle 45° until the mixing ratio arrives at 7.05% (p value <0.05).

The reasons for the influence of bottleneck angles on the walking speed are explored as follows. When the mixing ratio is relatively small (e.g. 2.35%), the position where the width of the crowd becomes narrow is similar (nearly 4m in x-coordinate) among angle 0°, 15° and 30°. However the taper position is further back in angle 45° because of the great contraction in the edge of the scene near the bottleneck, which can be found in Fig. 3 and Fig. 4(c). It is supposed that the phenomenon that the taper position of the trajectories is backward in angle 45° decreases the walking distance and makes the pedestrians move more directly towards the exit. Hence, the wheelchair user can move in relatively high speed in angle 45°due to this positive impact on participants' movement, while the differences among angle 0°, 15°, 30° are minor (p value >0.05). When the mixing ratio is relatively large (e.g. 7.05%), the existence of the large amount of wheelchair users with larger occupied area makes the crowd more clogging, especially in the scenario of angle 45° with less space near the bottleneck. Influenced by the negative impact, the average speed in angle 45° is not the fastest compared to other bottleneck angles at this mixing ratio.

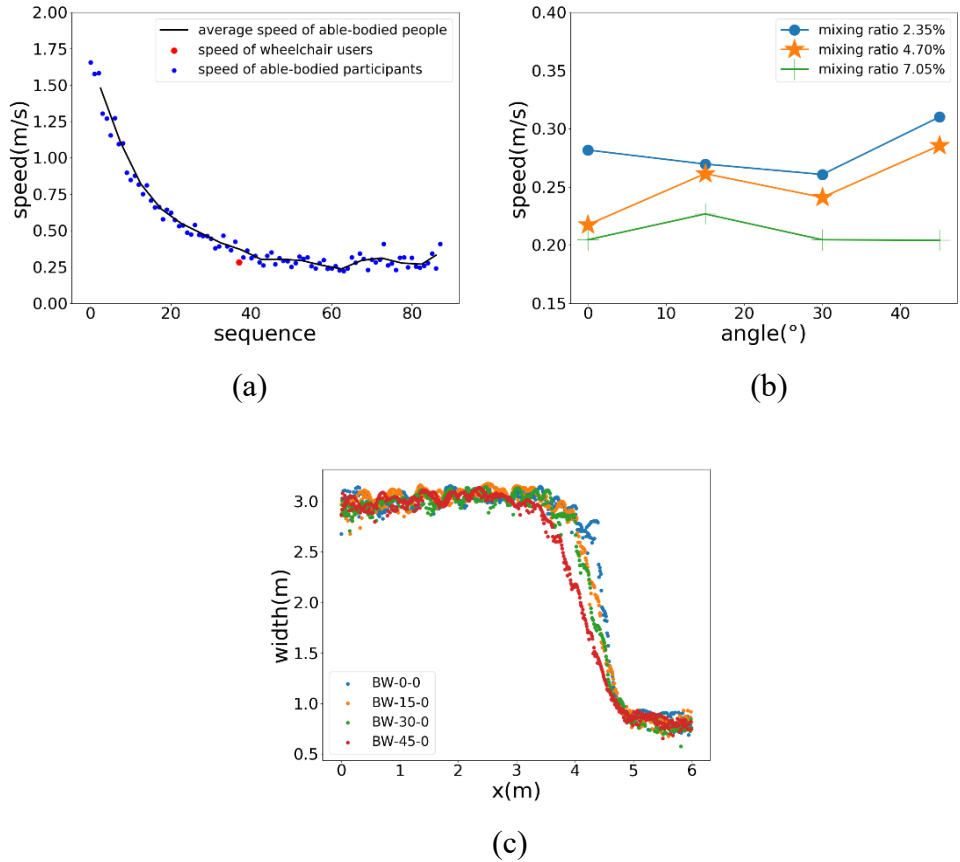

(a)                                  (b)

(c)

**Fig. 4.** (a) The relation between the average speeds of participants passing through the measurement area and their sequence of arrival. (Only scenario BW-0-3 is shown. More diagrams of other scenarios are shown in Appendix B). The red scatters mean the average speeds of wheelchair users and the blue ones represent the average speed of each able-bodied pedestrian. The black lines mean the average speeds of the group except the wheelchair users. (b) The diagram of the average speeds of the wheelchair users in different scenarios. (c) The relation between the width of the crowd and the corresponding x-coordinate position at the mixing ratio of 0% with different bottleneck angles. The results are calculated based on the trajectories of the crowd, which is shown in Fig. 3.

**Table 2** The average speeds of wheelchair users in different scenarios

| Scenario | BW-0-1 | BW-15-1 | BW-30-1 | BW-45-1 |
|---|---|---|---|---|
| Speed (m/s) | 0.282±0.152 | 0.270±0.126 | 0.261±0.134 | 0.310±0.097 |
| p value | / | 0.31 | 0.08 | 0.01 |
| Scenario | BW-0-2 | BW-15-2 | BW-30-2 | BW-45-2 |

| Speed (m/s) | 0.218±0.177 | 0.261±0.166 | 0.241±0.136 | 0.286±0.145 |
|---|---|---|---|---|
| p value | / | 0 | 0.01 | 0 |
| Scenario | BW-0-3 | BW-15-3 | BW-30-3 | BW-45-3 |
| Speed (m/s) | 0.205±0.122 | 0.227±0.135 | 0.205±0.124 | 0.204±0.138 |
| p value | / | 0 | 0.98 | 0.93 |

### 3.3 Escape time

The escape time of each trial, including individual time and total time, can indicate the movement efficiency. Here the individual escape time is defined as the period from one participant entering the starting line of the scenario to the departure of the bottleneck exit. The total escape time is the period from the first pedestrian entering the starting line of the scenario to the last one departing the bottleneck exit.

**(1) Individual escape time**

According to the definition mentioned above, the time period of each pedestrian passing through the bottleneck and the average value can be obtained. It needs to be noted that there is a significant difference between the speeds of the first 40 participants who arrived at the exit earlier and the other participants arriving later, which can be found in Fig. 4(a). Hence, instead of considering the whole group, which is used in [22], we calculate the average speeds of the latter group, namely the last 38 participants arriving at the exit. Besides, the variation ratio of a factor variable $VR_i$ which can reflect the gaps between itself $T_i$ and the reference $T_0$ (namely the factor value in angle 0° in this study) is calculated as:

$$VR_i = \frac{T_i - T_0}{T_0} \times 100\% \tag{2}$$

Hence, the negative value of $VR_i$ means that the value in the current scenario is smaller than that in angle 0° and vice versa.

Considering Fig 5(a), the individual escape time becomes larger with the mixing ratio in the way of quadratic function. The fitting parameters of the quadratic function are shown in Table 3. This reveals that the flow efficiency in the bottleneck becomes lower with the increase of mixing ratio from the perspective of individuals. The influence of the mixing ratio results from the lower speeds and larger occupied areas of wheelchair users. With the increase of wheelchair users, the crowd gets more clogging and the flow efficiency decreases.

Meanwhile, the influence of the bottleneck angles are shown in Fig. 5(b) and the variation ratios between the individual escape time in angle 0° and that in angle 15°,

30°, 45° respectively are calculated and shown in Fig. 5(c). Comparing to the individual escape time in the bottleneck of angle 0°, the value in angle 45° is smaller when the mixing ratio is 0% or 2.35% (the variation ratio are -3.6% and -2.8% respectively), while it becomes larger and the difference is minor at the mixing ratio of 4.70% and 7.05% (the variation ratio are -0.27% and 0.24% respectively). The individual escape time in angle 15° and 30° is larger than that in angle 0° and 45° in most cases. The results show that there is a significant advantage of utilizing the bottleneck of angle 45° at a low mixing ratio (lower than 2.35%) from the perspective of individual characteristics.

The causes for the effect of bottleneck angles are analyzed from the following aspects. We find that the participants take an extra walking distance compared to the linear distance between their initial position (namely on the starting line) and ending position (at the exit) because of the deflection of their motion direction. This phenomenon can also be observed from the tapered shape of the trajectories (see Fig 3). Therefore, two factors, namely the actual walking distance and average speed of the crowd, are utilized to reflect the flow efficiency. Considering the comprehensive impact of these two factors, the flow efficiency performs better when the actual walking distance of the crowd is smaller and the average speed is larger. We calculate the actual walking distance and average speeds of the pedestrians from the starting line of bottlenecks to the exit. Moreover, considering the apparent decline in the average speed of the first 40 participants arriving at the exit and the minor differences in the last 38 ones, the latter is considered both in the calculation of the two factors, which is the same as the calculation of individual escape time. The results of the actual walking distance and average speed are shown in Fig. 6.

According to Fig. 6(a), the actual walking distance in angle 15° and 30° is close to that in angle 0° (p value are 0.52 and 0.67, respectively and larger than 0.05) when the mixing ratio is down to zero. However, the actual walking distance in angle 45° is obviously smaller than that in angle 0° (p value <0.05). Fig. 6(b) indicates that the average speed in angle 45° is larger than that in other three angles. It is supposed that the position where the width of the crowd began to contract is similar among the scenarios of angle 0°, 15°, 30°, while the position is further back in angle 45° due to the great contraction in the edge of the scene near the bottleneck. These are shown in Fig. 3 and 4(c). According to the trajectories of the crowd, three points (two points of them represent the different positions where the width of crowd tapered and the other one represents the terminal point) form a triangle. The earlier the crowd starts to taper, the smaller walking distance will be taken by participants considering the three points and the principle that one side of a triangle is smaller than the sum of the other sides. Therefore, there is a positive effect on making participants moving towards the exit and a better flow efficiency in angle 45° because of its smaller actual walking distance and larger average speed.

When the mixing ratio is relatively larger, e.g. 7.05%, the average speeds decrease in angle 15°, 30° and 45°. We suppose that the contraction in the edge of these scenarios leads to the congestion of the crowd considering the large number of wheelchair users with larger occupied area moving towards the exit where the space is less. The comprehensive effect of the actual walking distance and average speed, which can be represented by individual escape time, illustrates that the individual flow efficiency in angle 45° falls to the value similar to that in angle 0° when the mixing ratio is larger than 4.70%. From another perspective, the finding that the advantage of flow efficiency in the bottleneck with an angle of 45° decreases can also be explained as follows. When the mixing ratio is relatively large, each wheelchair user had to change their paths of travel compared with the movement in small mixing ratio (e.g. only one wheelchair user) which is closed to a straight line centered in the stream and poses little disruption. Besides the change of travel path, the behavior of negotiating with or even yield to each other would also result in congestion and reduction in flow efficiency especially in the bottleneck with angle 45° considering its narrow edge and the large occupied area of wheelchairs.

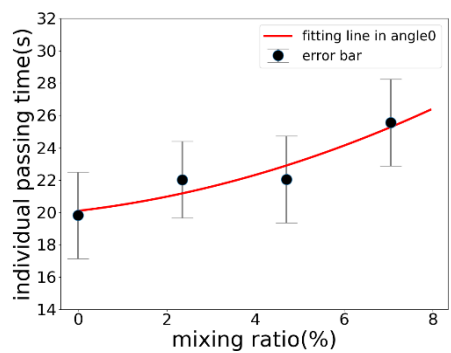

(a)

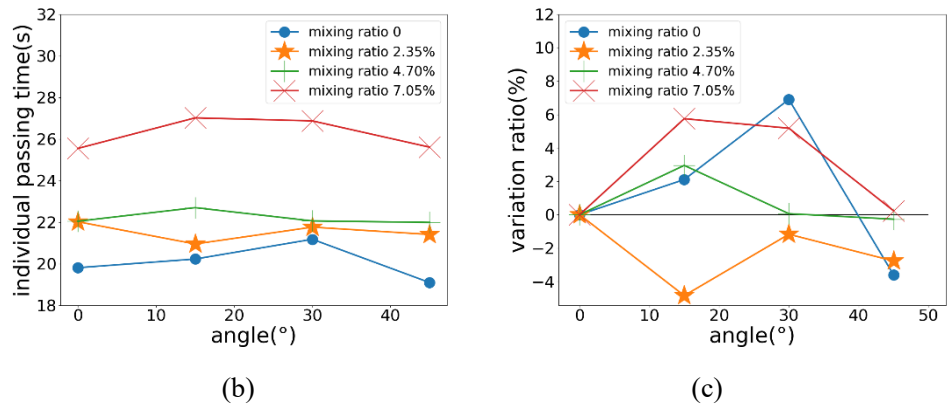

(b)                         (c)

**Fig. 5** (a) Diagrams of the relation between the individual escape time and the mixing ratio of wheelchair users and assistants at angle 0° (diagrams at other angles are shown in Appendix C). (b) The diagram of the relation between the average individual escape time and the bottleneck angles (c) The diagram of the variation ratio in each trial.

**Table 3** Parameters of the fitting curves between the individual passing time and mixing ratio

| Equation | $Y=ax^2+bx+c$ | | | |
|---|---|---|---|---|
| Value | a | b | c | R Square ($R^2$) |
| Angle 0° | 0.06 | 0.29 | 20.17 | 0.90 |
| Angle 15° | 0.16 | -0.21 | 20.31 | 0.99 |
| Angle 30° | 0.19 | -0.56 | 21.32 | 0.94 |
| Angle 45° | 0.09 | 0.12 | 20.04 | 0.96 |

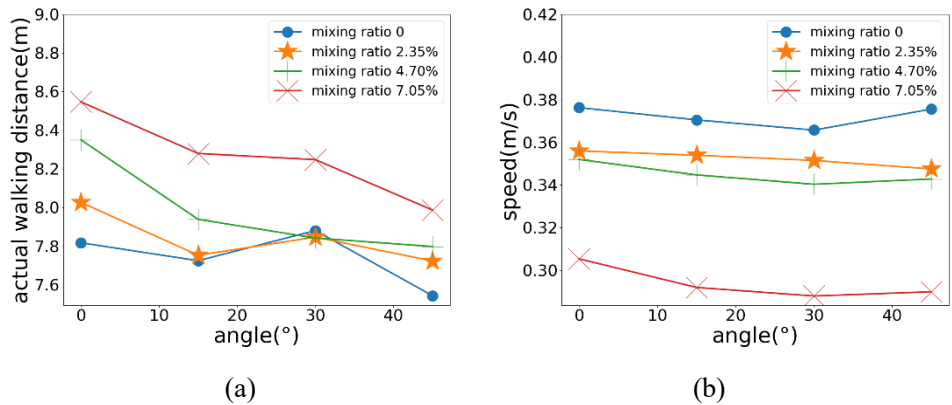

(a)                         (b)

**Fig. 6** The average of actual walking distance (a) the average speeds (b) of the last group in each scenario.

**(2) Total escape time**

In order to reflect the motion efficiency of the crowd at the group level, we utilize N-t diagrams to straightforwardly describe the moment when each participant leave the bottleneck exit and then calculate the total escape time of the crowd in each scenario for quantification. In this study, 91.7% (78 participants) of the whole crowd are considered to calculate the total escape time. Fig 7(a) describes the relation between the cumulated number of people leaving the bottleneck exit and the corresponding time needed for the departure. In Fig 7(a) we observe that the motion efficiency in angle 0° is similar to that in angle 15° and is slightly higher than that in angle 30°. Besides, the efficiency performs better in angle 45° than other three angles according to the slopes of these lines.

For the purpose of quantitative analysis of the motion efficiency, the total escape time of each trial is calculated (according to its definition or the points of intersection between the black and other lines in Fig. 7(a)) and plotted in Fig 7 (b) to (d). Fig 7(b) describes the relation between total escape time and mixing ratio of wheelchair users and assistants in the bottleneck of angle 0°. Meanwhile, the fitting curves are plotted and the value of fitting parameters as well as coefficient of determination (R square) are shown in Table 4. It is illustrated that the total escape time increases with the mixing ratio in the form of a quadratic function.

According to Fig 7(c) and 7(d), it can be found that the total escape time from 45° bottleneck is smaller than other three bottlenecks at the mixing ratio of 0 and 2.35% (the variation in angle 45° are -8.22% and -4.05% respectively), while the value in 45° bottleneck is close to that from 0° bottleneck at the mixing ratio of 4.70% and 7.05% (the variation in angle 45° are 0.72% and 0.28% respectively). The total escape time from 15° and 30° bottlenecks are larger than from 0° and 45° bottlenecks in most cases (except when the mixing ratio is 2.35%). The results indicate that the movement efficiency at the group level performs the best in the bottleneck with angle 45° when the mixing ratio is lower than 2.35%. However, the advantage of the bottleneck with angle 45° in movement efficiency disappears in the condition of relatively high mixing ratio (larger than 4.70%).

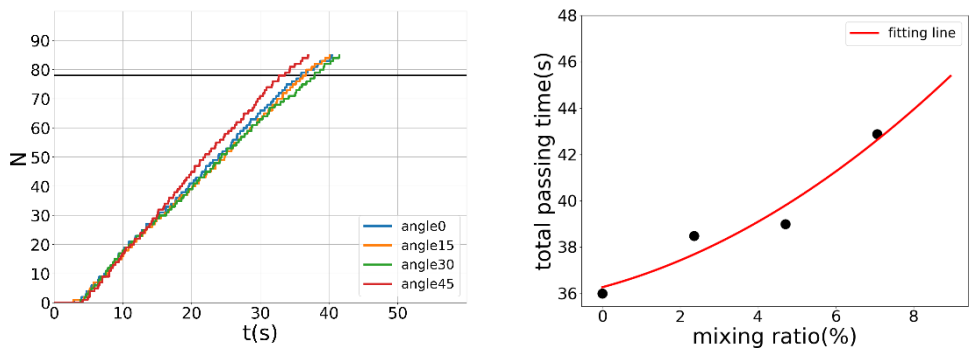

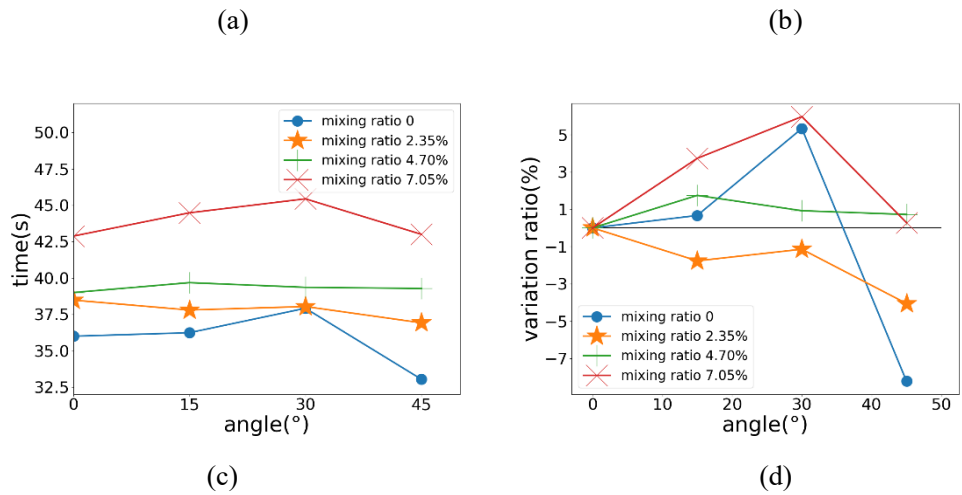

(a) (b)

(c) (d)

Fig. 7 (a) The relation between the cumulated number of people leaving the bottleneck exit and the corresponding time needed for the departure. Each line represents the relation in one type angle when the mixing ratio is 0. The relations in other mixing ratios are shown in appendix C. The points of intersection between the black and other lines represent the escape time for 78 participants at different angles. (b) Diagrams of the relation between total escape time and the mixing ratio of wheelchair users and assistants as well as their own fitting curves in the bottlenecks of angle 0° (diagrams in other angles are shown in appendix C) (c) Diagrams of the relation between total escape time and bottleneck angles. (d) The diagram of the variation ratio in each trial.

Table 4 Parameters of the fitting curves between the total escape time and mixing ratio

| Equation | $Y=ax^2+bx+c$ | | | |
|---|---|---|---|---|
| Value | a | b | c | R Square ($R^2$) |
| Angle 0° | 0.06 | 0.45 | 36.27 | 0.94 |
| Angle 15° | 0.14 | 0.10 | 36.37 | 0.99 |
| Angle 30° | 0.27 | -0.88 | 38.10 | 0.98 |
| Angle 45° | -0.01 | 1.42 | 31.84 | 0.99 |

Overall, the escape time including individual time and total time, indicates that the flow efficiency gets worse with the increase of the mixing ratio because of the increased number of wheelchair users with low ability of motion and large occupied area. Meanwhile, in this study the flow efficiency from the perspective of both individual and group performs best in angle 45° among all bottleneck angles when the mixing ratio of wheelchair users and assistants is lower than 2.35%. However, the advantages over the efficiency in angle 45° gradually disappear with the mixing ratio higher than 4.70%. The reasons can be analyzed from the actual walking distance and the average speed mentioned before.

## 3.4 Time-space relation

In order to describe the jams or stop-and-go waves in the crowd during participants' movement, the time-space relation is adopted and defined as the relation between the variation of motion in the direction towards the exit and the time required for the movement. Fig 8 describes the time-space relations in the measurement area of the bottleneck with angle 0°. The red lines represent that the individual velocity is lower than 0.1m/s and are considered in stop stage [25, 34], while the blue ones mean the velocity is higher than 0.1m/s. Take the case of the 0° bottleneck, it is found that the red lines appear more frequently with the increase of the mixing ratio. The phenomenon indicates that there are more participants with the speed lower than 0.1m/s (namely the participants considered in stop stage in this study) when the mixing ratio of wheelchair users and assistants increases. In order to describe and quantify the degree of this phenomenon of stopping and jam during crowd's movement, a factor R is introduced and defined as the ratio between the sums of each pedestrian's stopping time $t_i$ and the sum of each one's moving time $T_i$ in the measurement area:

$$R = \frac{\sum_{i=0}^{N} t_i}{\sum_{i=0}^{N} T_i} \times 100\% \tag{3}$$

According to Fig 9(a), it can be found that the factor R increases with the mixing ratio, which illustrates that the phenomenon of stopping and jams is more serious in the crowd. It is supposed that the increase of wheelchair users with smaller speed and larger occupied area has negative impact on the pedestrians' movement and exacerbates the congestion of the crowd.

With respect to the influence of the angles of bottlenecks, Fig. 9(a) and 9(b) show that the factor R in 45° bottleneck is smaller than in 0° bottleneck when the mixing ratio is lower than 4.70% (the variation ratio in 45° bottleneck are -9.91%, -4.88% and -24.81% respectively). However, when the mixing ratio is up to 7.05%, the factor R in 45° bottleneck is a little higher than in 0° bottleneck (the variation ratio in angle 45° is 4.91%). Besides, the values in 15°and 30° bottlenecks rise and fall at different mixing ratios compared to the value in angle 0°, while they are mostly higher than the value in 45° bottleneck. Based on the comparison of the factor R in different bottleneck angles, it is indicated that the degree of stopping and jams during the movement of the crowd is relatively smaller in 45° bottleneck all the time when the mixing ratio is lower than 4.70%. It is supposed that the great contraction of the boundary in 45° bottleneck makes the crowd began to taper early, which is shown in trajectories and lead to a positive impact on making participants move forward to the exit. However, the advantage of the angle 45° decreases when the mixing ratio is 7.05%. We conjecture that the contraction of the edge of the bottleneck in angle 45° leads to the clogging considering the large amount of wheelchair users with small walking speeds and larger covered area moving towards the exit.

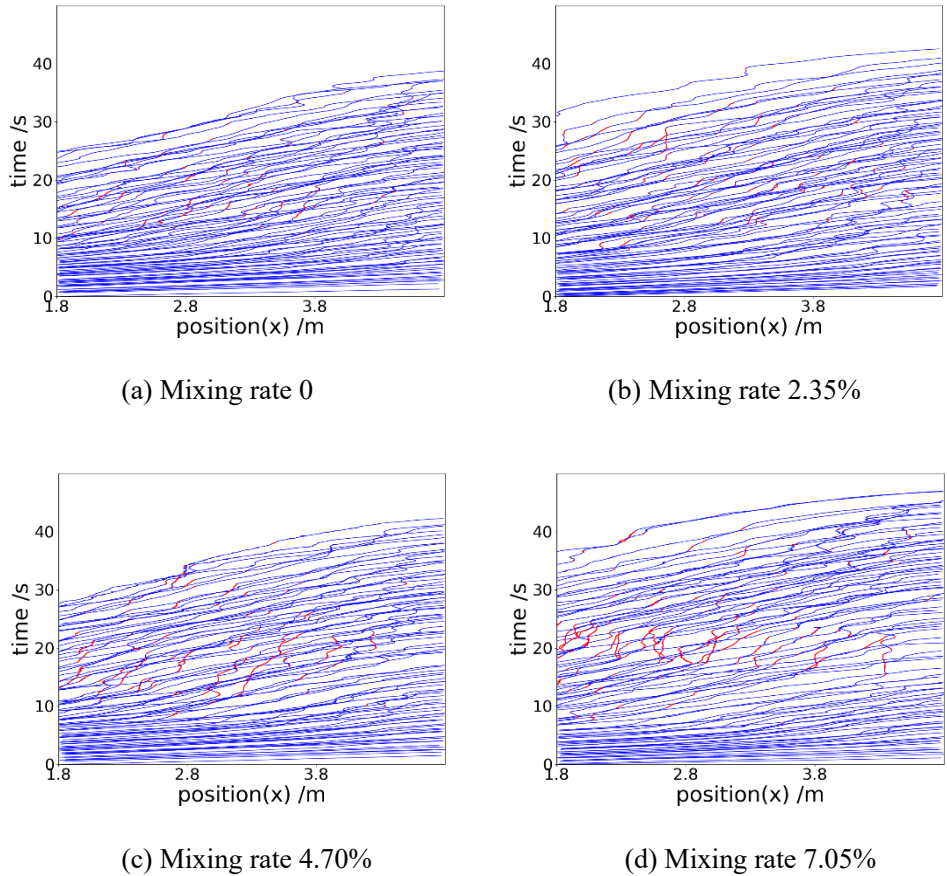

(a) Mixing rate 0          (b) Mixing rate 2.35%

(c) Mixing rate 4.70%        (d) Mixing rate 7.05%

Fig. 8 Diagrams of the relation between the exit-oriented movements and the time required for the motion under different mixing ratio of wheelchair users and assistants. (Only the relation in the bottleneck of angle 0 are given. See Appendix B for others in detail.)

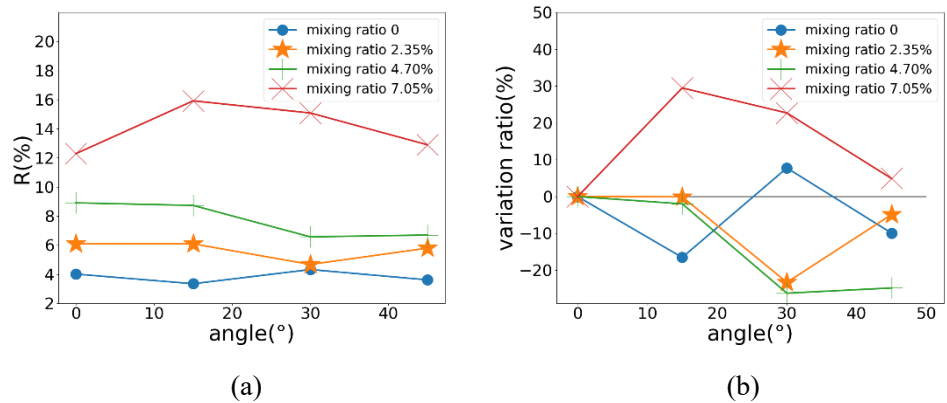

(a)          (b)

Fig. 9. (a) The relation between the factor R and the bottleneck angles in the circumstance of different mixing ratios. (b) The relation between the variation ratio of the factor R and the bottleneck angles under different mixing ratios.

**3.5 Time headway and its distribution**

For the purpose of reflecting the congestion of the crowd at the bottleneck entrance, time headway, namely time gap is utilized in this study. It is defined as the time interval between two consecutive pedestrians passing through the bottleneck entrance. The time headway is calculated according to the above definition.

The distribution of time headway is often utilized to help study the collective behavior and clogging of the crowd and power law distribution have successfully explained the phenomenon of jam and congestion. Therefore, we adopt power law distribution to investigate the clogging and jam of the crowd through bottlenecks in this study. Power laws, one of 'heavy-tailed' distribution, are one kind of probability distributions in the form of [35]:

$$p(x) \propto x^{-\alpha} \qquad (4)$$

Hence, the power law distributions obey a linear relation under the log-log scale coordinates, which can be shown as:

$$lnp(x) = -\alpha lnx + C \qquad (5)$$

It is effective to consider the complementary cumulative distribution function (CCDF for short) of a power law distributed variable in many studies [35, 36]. Hence, the CCDF of a random variable (the time headway) was calculated and fitted using power law distributions by the method described in [35]. The goodness-of-fit tests introduced in [36] were conducted and the results in each scenario passed the tests. The scaling parameter α was estimated from the fitting and shown in Fig 10(a), which can describe the severity of the clogging of pedestrian flows in accordance with the rule that the crowd becomes more congested with the decrease of the scaling parameter α.

The results shown in Fig 10 (a) indicate that the scaling parameter α becomes smaller with the increase of the mixing ratio. Hence, the congestion of the crowd at the entrance of the bottleneck becomes worse with the increase of the mixing ratio due to the slow movement and large area of wheelchair users. According to Fig. 10 (b), it is illustrated that the scaling parameter α in 45° bottleneck is smaller than other three angles (0°, 15°, 30°) when the mixing ratio is lower than 2.35%, which indicated less congestion in 45° bottleneck. However, the scaling parameter α for 45° gets close to that of 0° at the mixing ratio of 4.7% and 7.05% (the variation ratio in 45° bottleneck are 1.49% and -0.53% respectively), while the bottleneck of angle 15° and 30° both have a lower value of α all the time. Hence, the bottleneck of angle 45° tends to be a good choice at low mixing ratio (<2.35%) according to the degree of clogging because of the contraction of the boundary in angle 45° which is benefit for letting participants move forward to the exit.

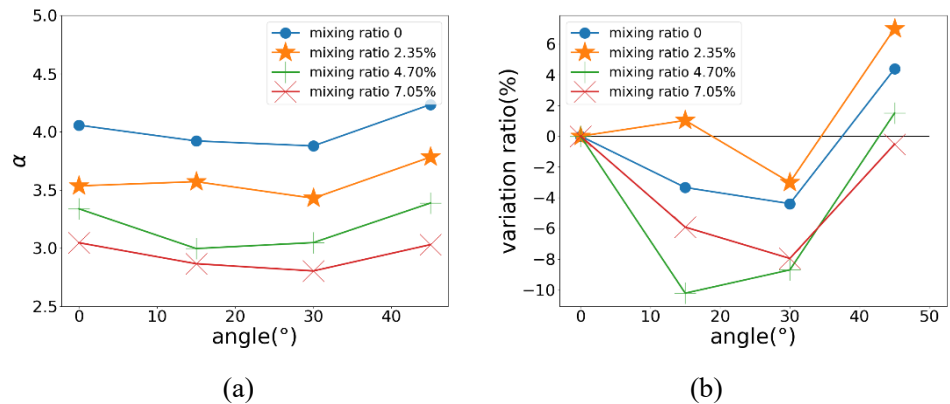

(a)　　　　　　　　　　　　(b)

Fig. 10 (a) The diagram of the relation between scaling parameter α and the bottleneck angles in the circumstance of different mixing ratios. (b) The diagram of the variation ratios.

## 4. Conclusions

It is of great importance to study the characteristics of the movement of pedestrian flows including wheelchair users as the increase on the number of the elderly and disabled in China nowadays. For the purpose of improving the efficiency and safety of pedestrians' motion and evacuation in bottleneck areas where congestion and stampede often occurred, controlled pedestrian experiments in the bottleneck including wheelchair users were conducted in this study. Moreover, we arranged different number of wheelchair users with assistants to stand evenly in a row in the middle of the crowd and changed the outer margin of the bottleneck in the controlled experiments. The factors such as speed, escape time, stopping time and time headway are analyzed to explore the impact of wheelchair user mixing ratios and bottleneck angles on the sporting ability, efficiency and congestion of pedestrian flows. According to the results of these experiments, conclusions are summarized as follows based on the arrangement of the wheelchair users and the bottleneck in this study:

(1) The speeds of wheelchair users, which can reflect the movement abilities of them, get lower with the increase of the mixing ratio. Besides, the average speeds of wheelchair users in the bottleneck is the fastest in 45° bottleneck (0.310±0.097m/s) until the mixing ratio arrives at 7.05%.

(2) The presence of wheelchair users leads to worse efficiency of traffic and congestion in the bottleneck and the degree of them increases with the mixing ratio of wheelchair users and assistants. Moreover, the variation of the escape time on the mixing ratio coincides with the quadratic function in this study.

(3) The bottleneck of angle 45° reflects a better efficiency of movement and less degree of stopping and congestion (2.8%~9.9% according to the variation ratios of the factors in Section 3) in the pedestrian flows under the circumstance of low mixing ratio (<2.35%) compared to other three angles (0°, 15°, 30°). However, the advantages of the angle 45° disappear in the condition of high mixing ratio as the factors are close to that in angle 0°.

The findings in this study is meaningful for the guidance of adding specific facilities for evacuation in the bottlenecks considering the presence of wheelchair users. On the one hand, the relation between the traffic efficiency or clogging and the mixing ratio indicates the worse evacuation in the presence of large number of wheelchair users and reminds people to pay more attention on the safety in such situation. On the other hand, it is effective to add a funnel shaped zone with angle 45° under the circumstance of no wheelchair users or the mixing ratio relatively small (2.35% in this study). However, in this study we only recruited limited number of wheelchair users who were mimicked by able-bodied participants and arranged to stand evenly in a row in the middle of the crowd through the specific dimension of the bottleneck. Our study only contains four different angles of bottleneck without repeated experiments. The explanation for the impacts of bottleneck angles needs further experimental verification. Further researches including different dimensions of the bottlenecks and the increase of the mixing ratio as well as different distribution of wheelchair users are necessary in the following study.

**Acknowledgements**

The authors acknowledge the foundation support from the National Natural Science Foundation of China (Grant No. 71704168, U1933105), from Anhui Provincial Natural Science Foundation (Grant No.1808085MG217) and the Fundamental Research Funds for the Central Universities (Grant No. WK2320000040).

## Appendix A

Diagrams of the trajectories of each scenario. The grey lines represent the trajectories of the able-bodied participants, while the red one means the trajectories of the wheelchair user and the green one means that of the assistant.

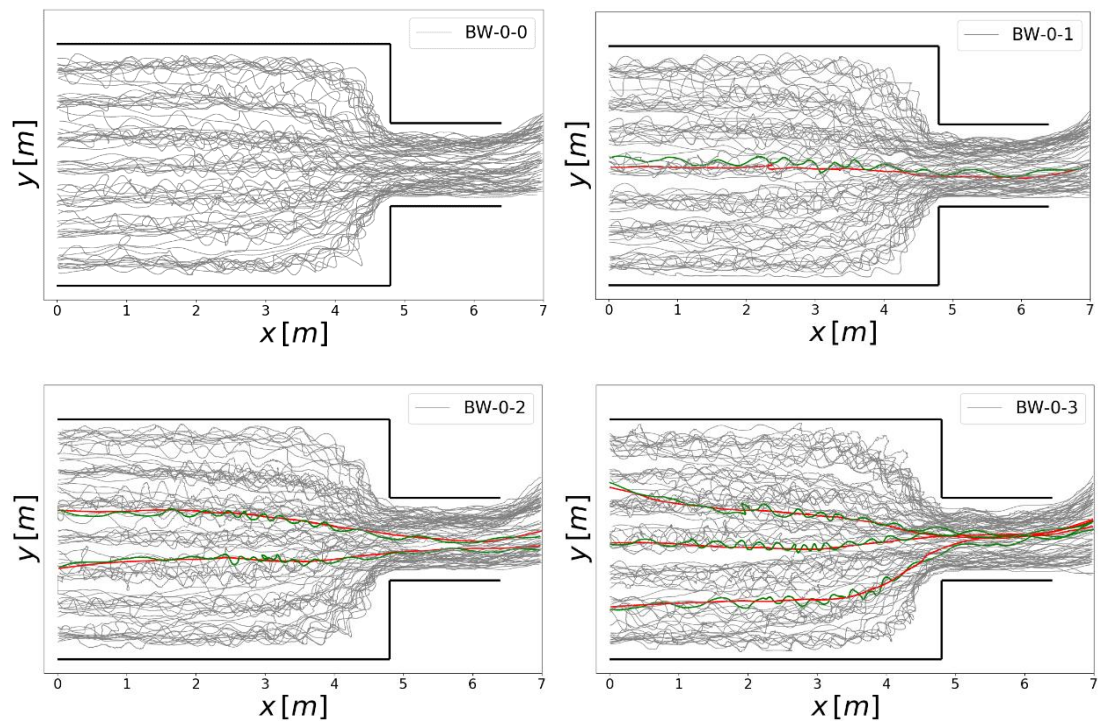

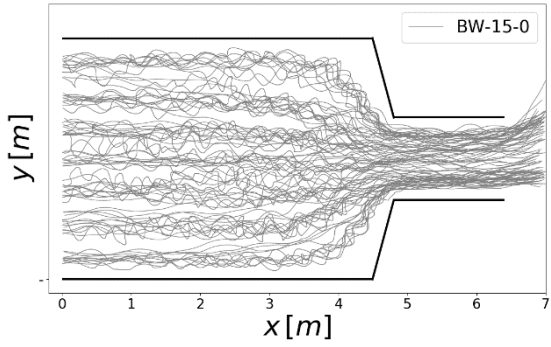
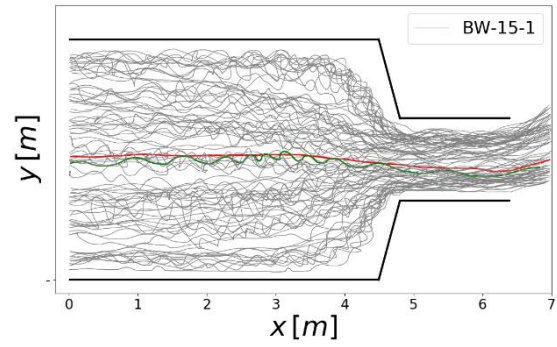
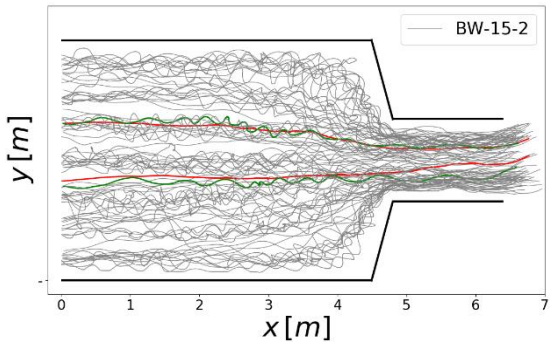
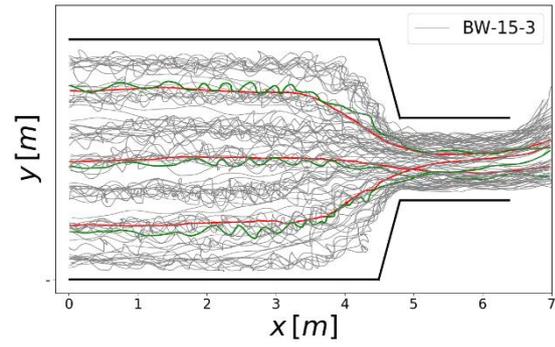
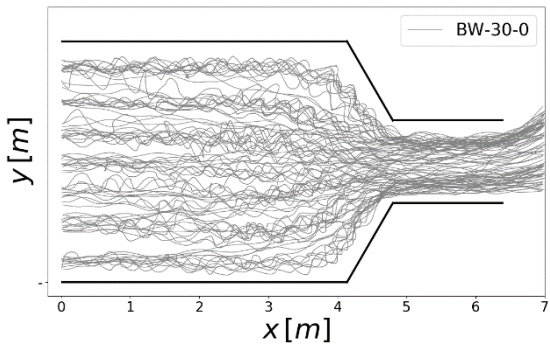
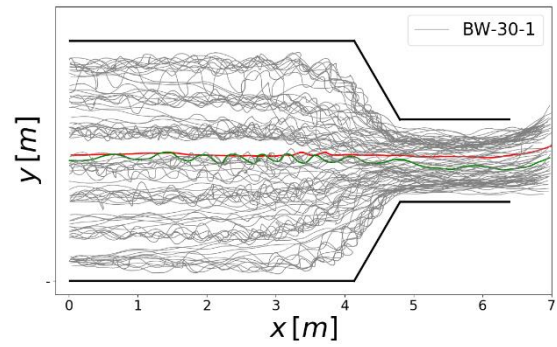
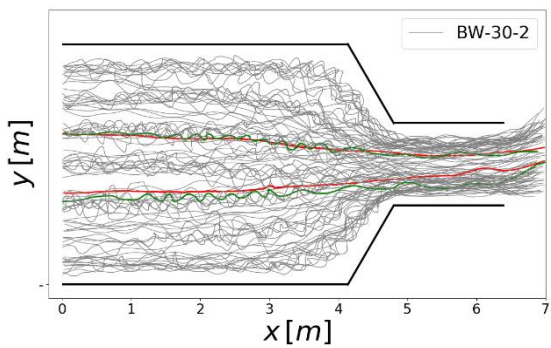
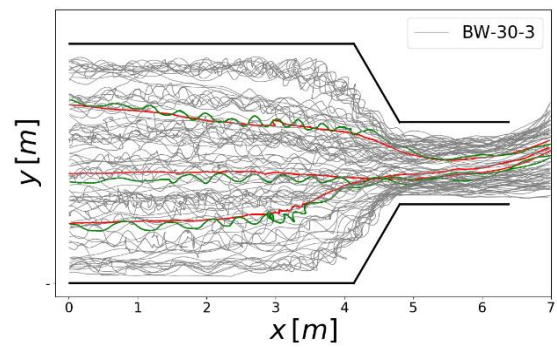
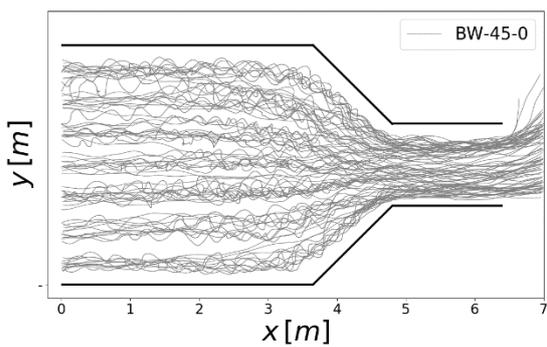
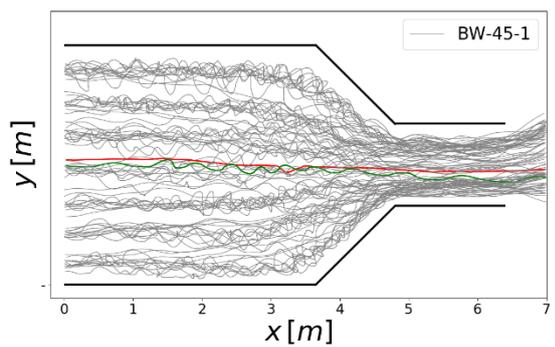

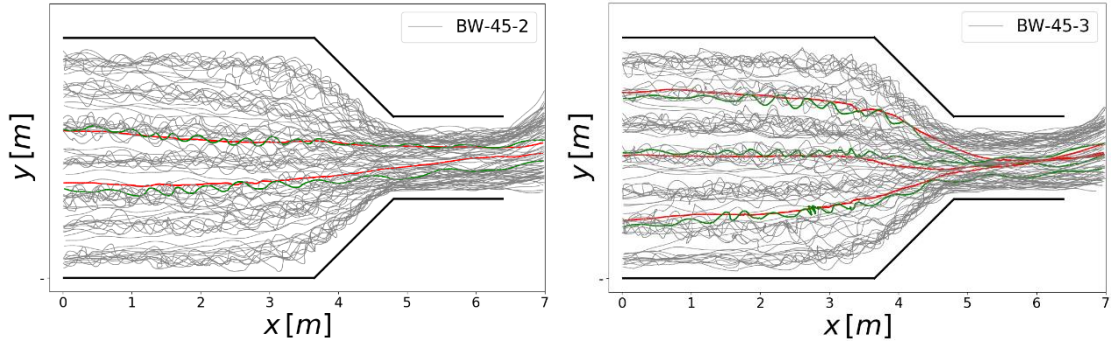

## Appendix B

Diagrams of the relation between the average speeds of participants passing through the measurement area and their sequence of arrival. The red scatters mean the average speeds of wheelchair users and the blue ones represent the average speed of each able-bodied pedestrian. The black lines mean the average speeds of the group except the wheelchair users.

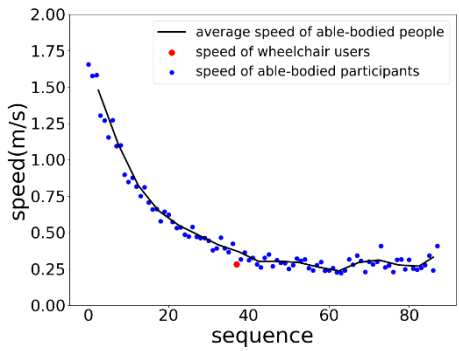

BW-0-1

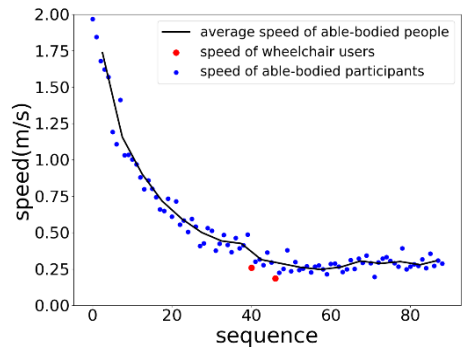

BW-0-2

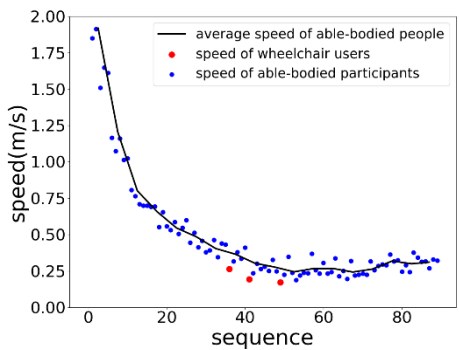

BW-0-3

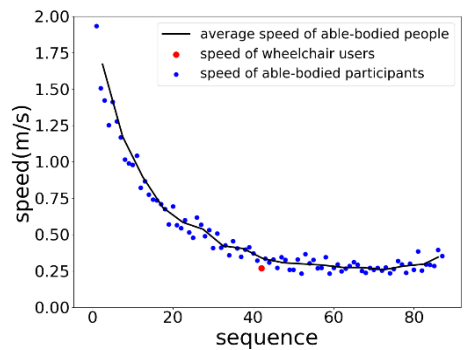

BW-15-1

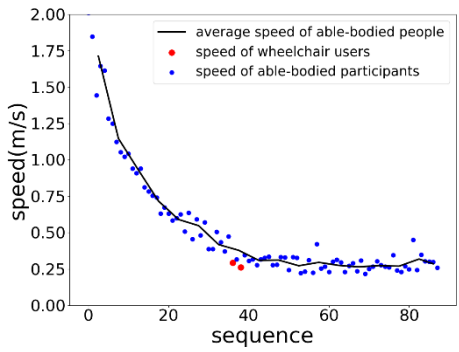

BW-15-2

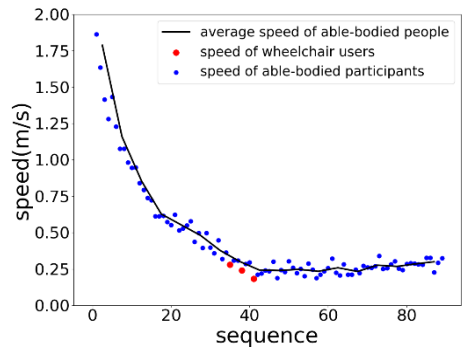

BW-15-3

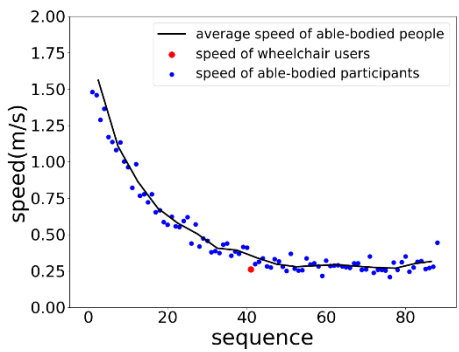

BW-30-1

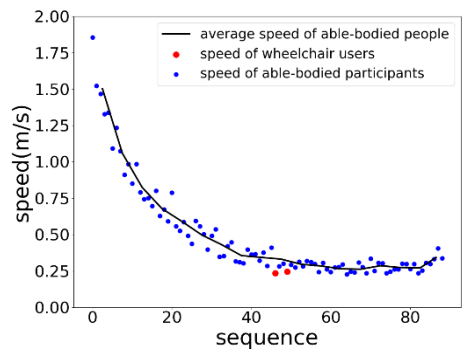

BW-30-2

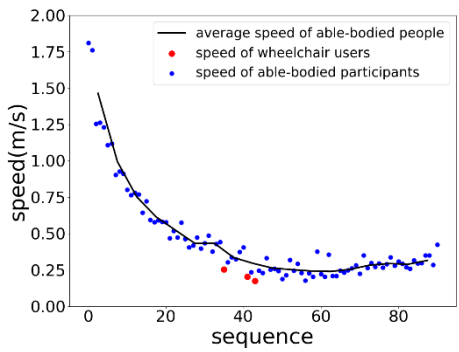

BW-30-3

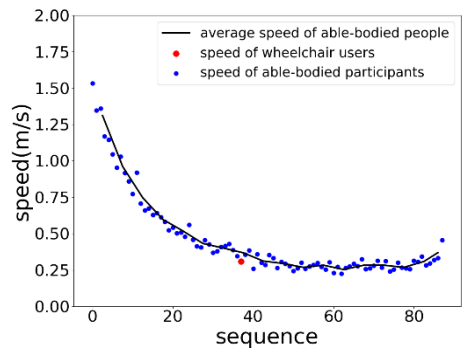

BW-45-1

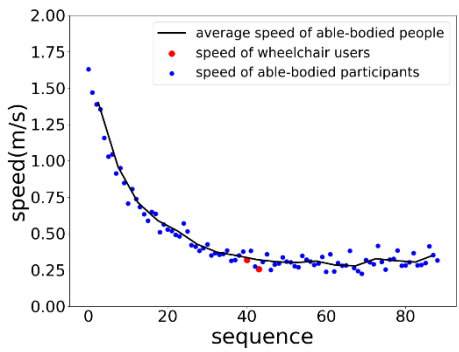

BW-45-2

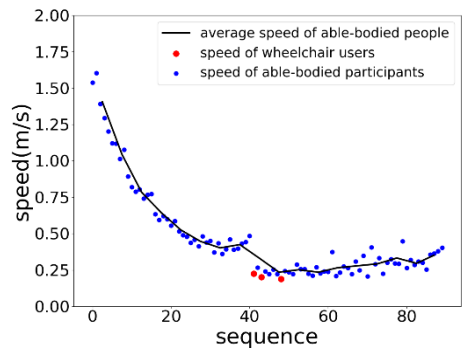

BW-45-3

**Appendix C**

(1) Diagrams of the relation between the individual escape time and the mixing ratio of wheelchair users and assistants at different bottleneck angles.

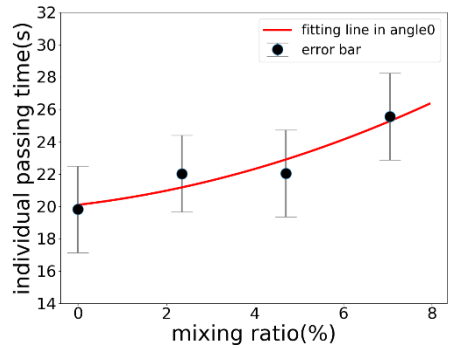

Angle 0°

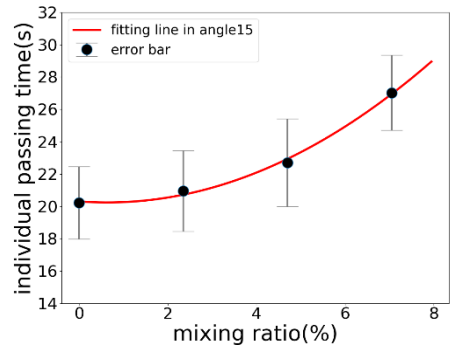

Angle 15°

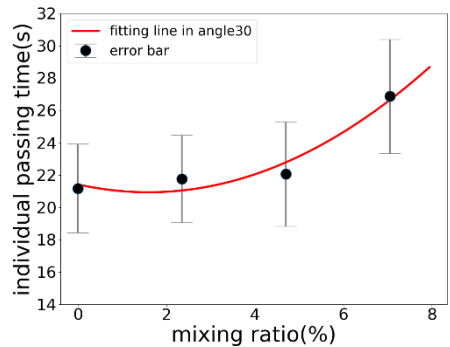

Angle 30°

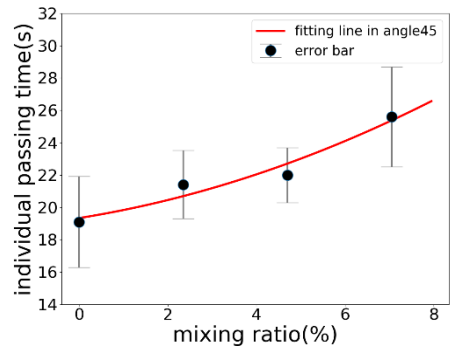

Angle 45°

(2) Diagrams of the relation between the cumulated number of people leaving the bottleneck exit and the corresponding time needed for the departure.

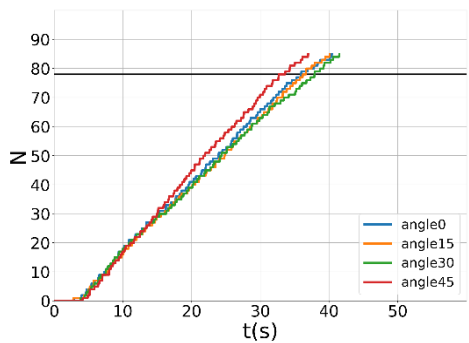

Mixing ratio 0

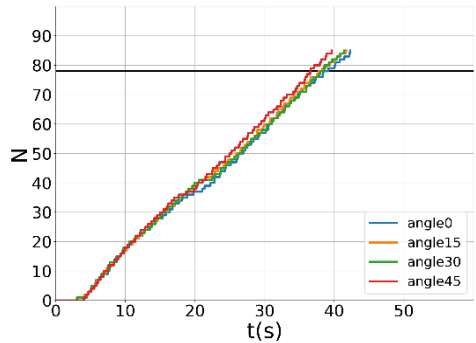

Mixing ratio 2.35%

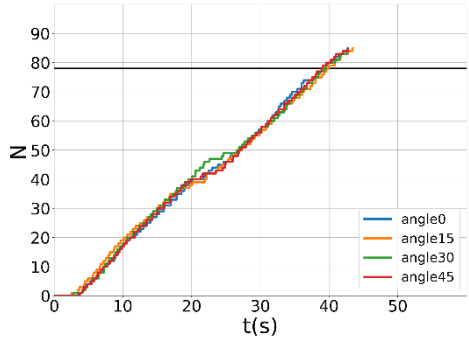
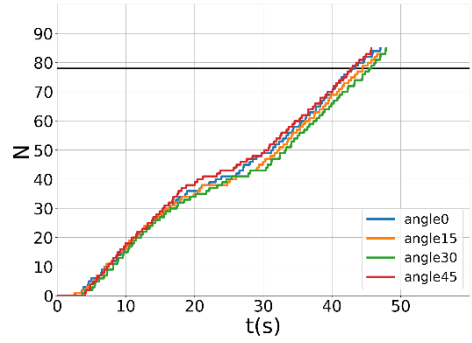

Mixing ratio 4.70%          Mixing ratio 7.05%

(3) Diagrams of the relation between total escape time and the mixing ratio of wheelchair users and assistants at different bottleneck angles

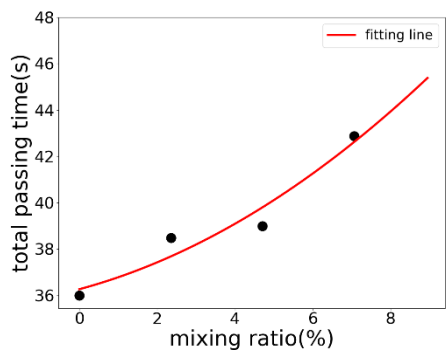
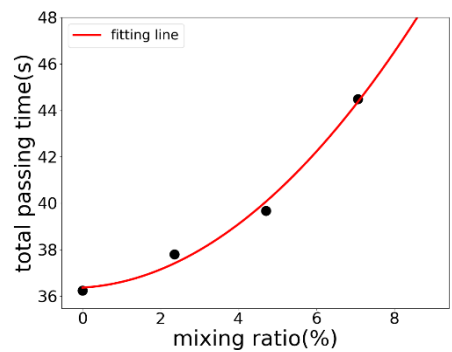

Angle 0°          Angle 15°

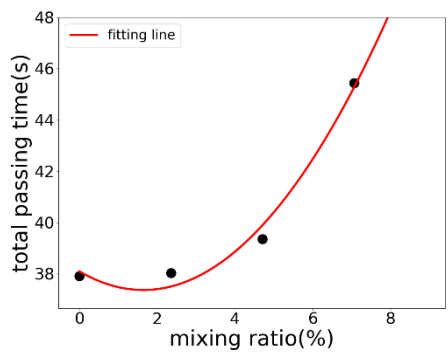
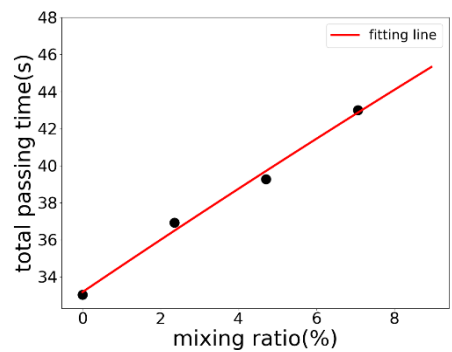

Angle 30°          Angle 45°

**Appendix D**

Diagrams of the relation between the exit-oriented movements and the time required for the motion under different mixing ratio of wheelchair users and assistants.

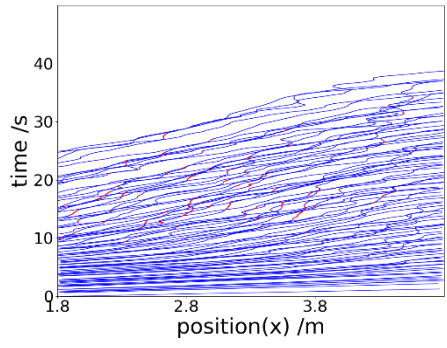

BW-0-0

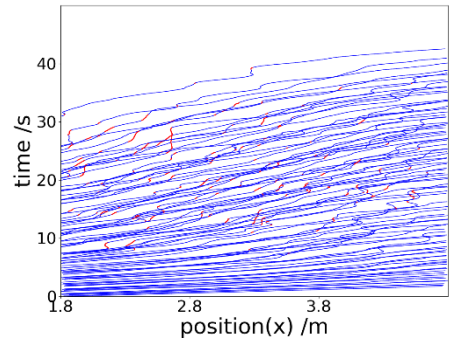

BW-0-1

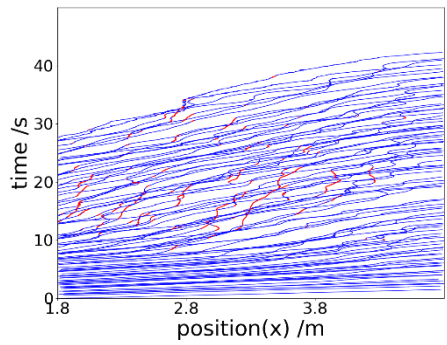

BW-0-2

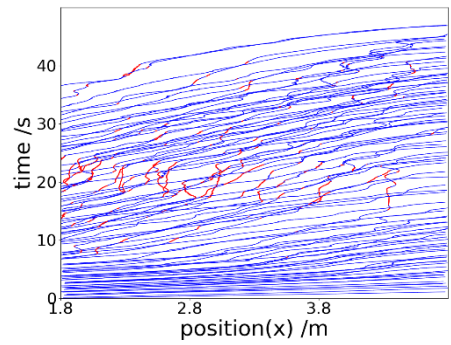

BW-0-3

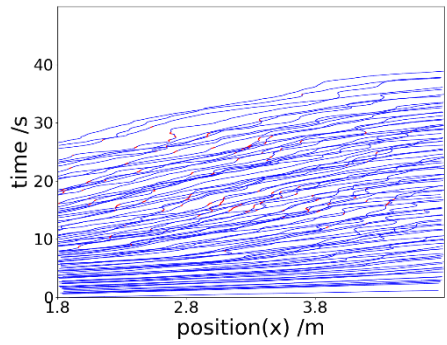

BW-15-0

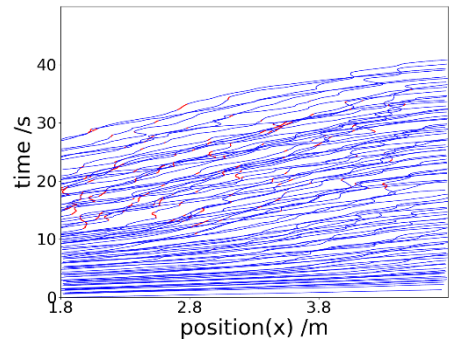

BW-15-1

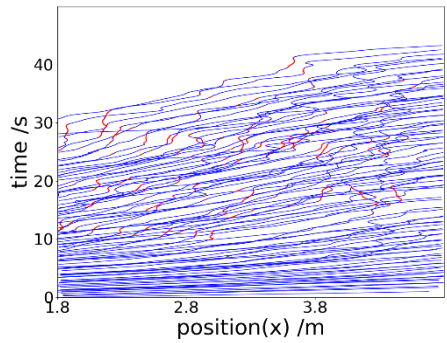

BW-15-2

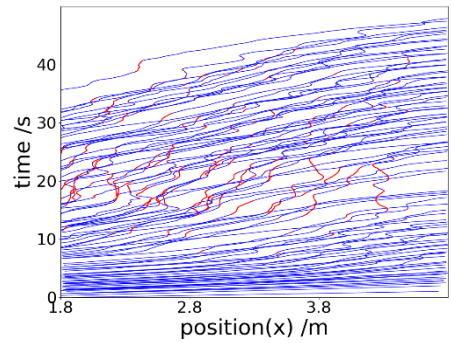

BW-15-3

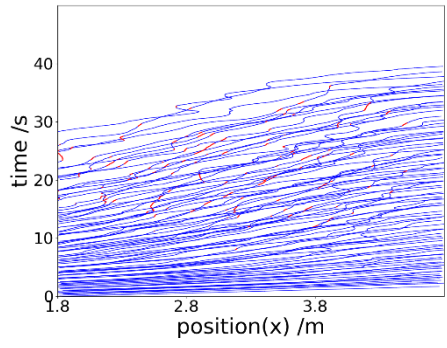
BW-30-0

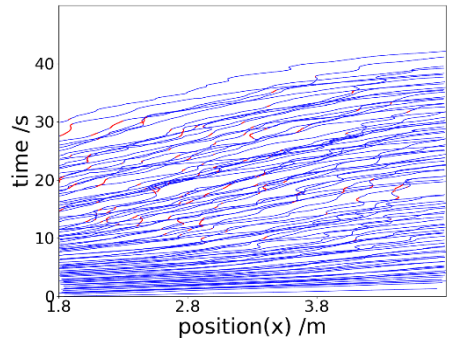
BW-30-1

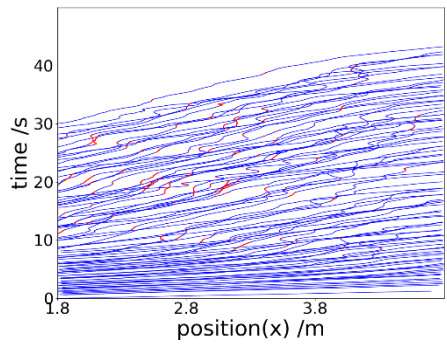
BW-30-2

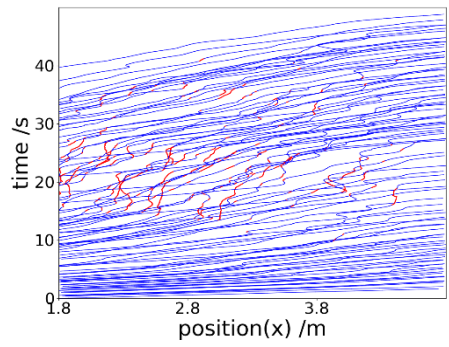
BW-30-3

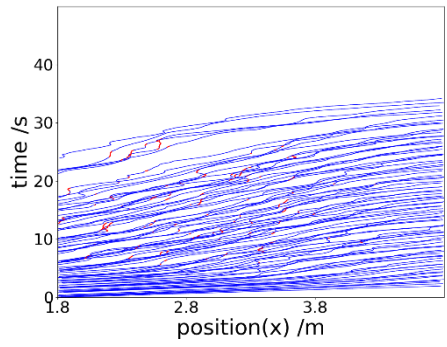
BW-45-0

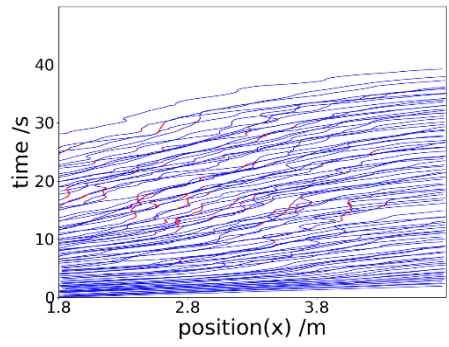
BW-45-1

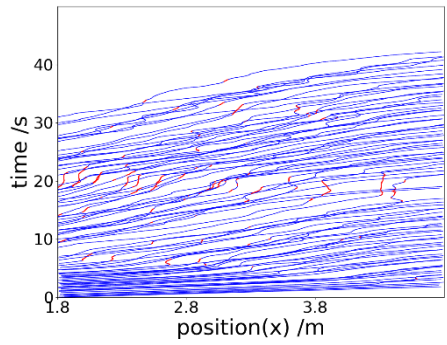
BW-45-2

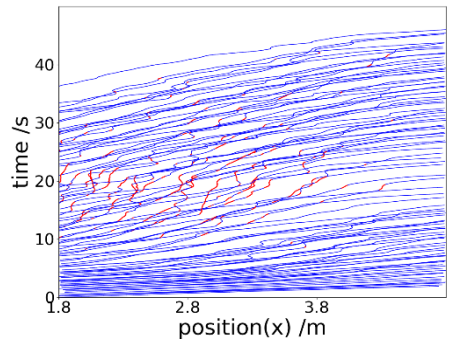
BW-45-3